# Continuum Model For Couette-Poiseuille Flow In A Drag Molecular Pump

P. A. Skovorodko

*Institute of Thermophysics, 630090, Novosibirsk, Russia*

## ABSTRACT

A continuum one-dimensional model of the plane Couette-Poiseuille flow is developed to describe the pressure distribution in a drag stage of molecular pump of either the Gaede or Holweck type. In spite of its simplicity and approximate nature the model provides a good qualitative representation of the drag pump operation in the whole range of the regimes from the continuum to free molecular ones.

## Model And Results

The steady Couette-Poiseuille flow in a long plane channel between two plates with equal temperatures $T$, one of which is moving with the velocity $w$ is described in the frames of the boundary layer equation $\rho u u'_x + p'_x = \mu u''_{yy}$ assuming the flow to be isothermal. The parameters $a$, $b$ and $c$ of the assumed parabolic velocity profile $u = ay^2 + by + c$ are derived from the value of mean gas velocity $\bar{u}$ and two boundary conditions on the plate surfaces counting for slip effects $\Delta u = \frac{(2-\sigma)}{\sigma}\frac{c_0 \mu}{p} u'_y = \lambda u'_y$, where $c_0 = \left[\frac{\pi RT}{2}\right]^{1/2}$.

The pressure distribution is mainly determined by the parameter $a$, which is equal to $3(w - 2\bar{u})/(6\lambda\delta + \delta^2)$. The parameters $b$ and $c$ are expressed in similar form. The resulting equation relatively $\bar{u}(x)$ may be obtained by integration of the governing equation across the channel. The continuity equation $g = \rho \bar{u} \delta(x)$ together with the equation of state $p = \rho RT$ close the problem. Similar approach was applied earlier for describing a critical flow through the plane channel [1], though the slip velocity was not inserted in the velocity profile definition.

The calculations show that for conditions typical of modern drag pumps the contribution of inertial term is small. Neglecting this term simplifies the governing equation to the form $p'_x = 2a\mu$ that for some shapes of the channel allows analytical solution. Thus for the channel of constant cross section $\delta$ and $L$ in length the relation between inlet ($p_1$) and outlet ($p_2$) pressures has the form

$$L = \frac{\delta^2}{6\mu w}\left[p_2 - p_1 + \frac{2}{\delta}\left(\frac{3(2-\sigma)}{\sigma}\mu c_0 + \frac{gRT}{w}\right)\ln\left|\frac{p_2[w - 2\bar{u}(p_2)]}{p_1[w - 2\bar{u}(p_1)]}\right|\right]. \qquad (1)$$

The relation (1) describes the distribution of pressure $p_1(L)$ inside the channel treating the length $L$ as the variable. This relation predicts reasonable results for all the regimes from the continuum to free molecular ones. From the analysis of this relation and the results for tapered channel ($\delta'_x < 0$) the following conclusions may be derived:

- In continuum regime the pressure difference $\Delta p_c = p_2 - p_1$ became constant and equal to $6\mu w L/\delta^2$ [2].
- In the free molecular regime the compression ratio $p_2/p_1$ became constant and equal to $\exp[\sigma w L/(2-\sigma)c_0 \delta]$.

- The terminal compression ratio in the free molecular regime is squared of that realized at $p_2 = \Delta p_c / 2$.
- The maximum throughput ($p_1 = p_2$) of the pump is determined by the condition $\bar{u} = w/2$ [2].
- An effective cross section of the tapered channel with $\delta'_x = const$ is equal to $(\delta_1 \delta_2)^{1/2}$ for the continuum regime and $(\delta_1 - \delta_2)/\ln(\delta_1/\delta_2)$ for free molecular one.
- The mean velocity at maximum throughput ($p_1 = p_2$) for the tapered channel may noticeably exceeds the value $w/2$ due to the pressure gradients inside the channel.

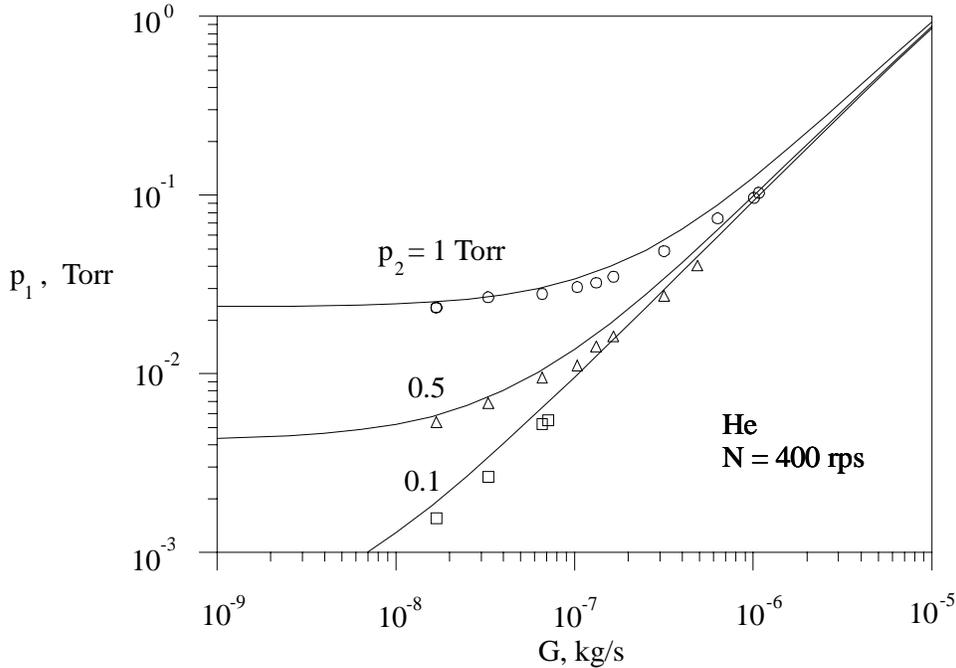

**FIGURE 1.** Inlet pressure versus gas flow rate at several outlet pressures.

The model can provide not only a qualitative description of the drag pump operation but also a quantitative one. For this purpose the accommodation coefficient $\sigma$ should be treated as a fitting parameter which value is to be derived from the compression ratio of the pump in the free molecular regime. Figure 1 illustrates the description of experimental results [3] (symbols) obtained for the Holweck pump. The predictions of the model shown by solid lines are obtained for $\sigma = 0.6$.

## Conclusion

In spite of its simplicity and approximate nature the developed model provides a good qualitative representation of the pumping process in the drag molecular pump in the whole range of the regimes from the continuum to free molecular ones. The possibility of quantitative description of the experimental results indicates that the model may be applied to optimize the design of the real pump, though the main purpose of the model consists in better understanding the basic features of the flow.